# Online Algorithms for a Generalized Parallel Machine Scheduling Problem

István SZALKAI[1,2], György DÓSA[1,3]

[1] Department of Mathematics, University of Pannonia, Veszprém, Hungary
[2] szalkai@almos.uni-pannon.hu
[3] dosagy@almos.uni-pannon.hu

January 25, 2015.

**Abstract:**  We consider different *online* algorithms for a generalized scheduling problem for *parallel machines*, described in details in the first section. This problem is the generalization of the classical parallel machine scheduling problem, when the makespan is minimized; in that case each job contains only one task. On the other hand, the problem in consideration is still a special version of the workflow scheduling problem.
  We present several heuristic algorithms and compare them by computer tests.

**Keywords:**  scheduling, parallel machines, online algorithm.

## 1 The generalized Parallel Machine Scheduling problem

We are given a list of (types of) **tasks $T_1,...,T_t$** , and another list of **machines $M_1,...,M_m$** , and the table **I[$i,j$]** which shows the required **time** what machine $M_i$ needs to solve the task $T_j$  ($1 \leq i \leq m$, $1 \leq j \leq t$).  I[$i,j$]≠0 is assumed, but I[$i,j$]<0 indicates that $M_i$ is unable to solve $T_j$ .

We receive online the list of  **jobs $J_1,...,J_L$** , where any job  $J_\ell$  ($1 \leq \ell \leq L$) contains several tasks from the list $T_1,...,T_t$ ,  in blocks (see /0a/ below).  The number of the jobs, L , and the number of *tasks* from which $J_\ell$ are build up, are unknown in advance. We have to schedule all the tasks of the jobs $J_\ell$  ($1 \leq \ell \leq L$) to the machines, fulfilling the following requirements.

/0a/   $J_\ell$   contains of  $f_\ell$  many **blocks**

$$J_\ell = (B_{\ell,1} ,..., B_{\ell,f_\ell}) \qquad (1)$$

/0b/   where any block $B_{\ell,\phi}$  ($\phi \leq f_\ell$)  contains some tasks

$$B_{\ell,\phi} = (T_{\ell,\phi,1} ,..., T_{\ell,\phi,K_\phi}) \qquad (2)$$

$K_\phi$ is called the **length** of $B_{\ell,\phi}$, assuming $1 \leq K_\phi$. The case $f_\ell=1$ is allowed, too. The **size** of the job $J_\ell$ is clearly

$$\text{size}(J_\ell) = \sum_{\Phi=1}^{f_\ell} K_\Phi \qquad (3)$$

/0c/   Since $T_{\ell,\phi,k}$ are members of the list $T_1,...,T_t$ we can define the function $\varphi$ as

$$\varphi(\ell,\phi,k) = \tau \quad \textit{if and only if} \quad T_{\ell,\phi,k}=T_\tau \qquad (4)$$

/0d/ any machine $M_i$ can solve any task, starting at any time, assuming $I[i,\tau]>0$.

/1a/   For any $\ell \leq L$ and $\phi \leq f_\ell$ the tasks of the block $B_{\ell,\phi}$ (see /0b/) can be solved (started) at any time, independently from each other (using any machine), however the blocks $B_{\ell,2},..., B_{\ell,\phi},..., B_{\ell,\phi+1}$ must wait for finishing the previous ones:

/1b/  *any task* of $B_{\ell,\phi+1}$ may start only when each task of $B_{\ell,\phi}$ has already been finished.

In other words: for any $\ell \leq L$, $\phi_1 < \phi_2 \leq f_\ell$ and $k_1 \leq K_{\phi_1}$, $k_2 \leq K_{\phi_2}$ the task $T_{\ell,\phi_2,k_2}$ can be started *only after* the task $T_{\ell,\phi_1,k_1}$ has been finished.

We underline that /1b/ refers to blocks of the *same job* $J_\ell$:

/1c/   we have *no* resctriction at all for the starting times of the tasks of the job $J_{\ell_1}$ when comparing to $J_{\ell_1}$, for $\ell_1 \neq \ell_2 \leq L$.

This means, especially, that

/1d/   all *tasks* (in the first block) of any job may be started even at time 0,

/2/    each machine $M_i$ in every time may work on *at most one* task $T_{\ell,\phi,k}$, and $M_i$ can not stop until it finishes the current task,

/3/    **scheduling** the task $T_{\ell,\phi,k}$ to the machine $M_i$ means, that we choose a (positive) number *d* such that $M_i$ is able to solve the task $T_\tau$ in the interval $(d, d+\text{Ido}[i,\tau])$, assuming $\tau=\varphi(\ell,\phi,k)$ and fulfilling /2/.

**The goal** is to finish all jobs as early as possible:

/4/    We have to finish all tasks of the jobs $J_1,...,J_L$ such that each machine finishes all its tasks until the time *I*, satisfying /0a/ through /3/ and *I* is minimal.

Let us observe, that /1b/ is the hardest part of our algorithmic problem. Assumption /1a/ is void when $K_\phi=1$. The other extreme case is when $f_\ell=1$, in this case /1a/-/1c/ imply that all tasks of $J_\ell$ can be scheduled in arbitrary manner (fulfilling /0a/-/0c/ and /2/ of course).

Naturally all task $T_{\ell,\phi,k}$ we have to schedule, is a member of exactly one block.

The problem could be described and solved using Integer Linear Programing,too,but the number of variables and equations grows exponentially ([2],[3]).

## 2 The algorithms

Recall, that the input is (in fact) the *one-dimensional sequence* of the tasks

$$T_{1,1,1}, \ldots, T_{\ell,\phi,k}, \ldots, T_{L,\phi fL,KfL} \tag{5}$$

where the terms $T_{\ell,\phi\ell,k}$ are for $1\leq\ell\leq L$, $1\leq\phi\leq f_\ell$ and $1\leq k\leq K_\phi$. Of course the input contains also the *delimiters* for determining the jobs and blocks in (4) according to /0a/ and /0b/. Single tasks without delimiters are considered one-element blocks: $B_{\ell,\phi}=(T_{\ell,\phi,1})$, i.e. $K_\phi=1$.

Our **problem** is the *online scheduling*: we have to schedule each task $T_{\ell,\phi,k}$ *immediately* after reading it:

/5/ receiving $T_{\ell,\phi,k}$ we have to choose $M_i$ and the starting time $d$ such that $M_i$ can solve $T_{\ell,\phi,k}$ in the interval $(d, d+\text{Ido}[i,\tau])$ without a break, where $\tau=\varphi(\ell,\phi,k)$.

Clearly, when deciding /5/, we have no information on the further tasks or on the length of the job or block we are working on, even not the number of jobs. Our scheduling in /5/ can not be altered later, of course.
(One illustrative example can be found in the 3rd section.)

In our research we implemented, tested and compared the following variants (numberings refer to our developing):

**Variant )3(** : Give the next task $T_{\ell,\phi,k}$ to the machine $M_i$ **if**

$M_i$ can *finish* $T_{\ell,\phi,k}$ the soonest,

i.e. $d+\text{Ido}[i,\tau]$ has the possible smallest value, where $\tau=\varphi(\ell,\phi,k)$ and $d$ is the starting time.

**Variant )4(** : Give the next task $T_{\ell,\phi,k}$ to the machine $M_i$ **if**

$M_i$ can *start* $T_{\ell,\phi,k}$ the soonest,

i.e. $d+\text{Ido}[i,\tau]$ has the possible smallest value.

**Variant )5(** : Give the next task $T_{\ell,\phi,k}$ to the machine $M_i$ **if**

$M_i$ can *solve* $T_{\ell,\phi,k}$ in the shortest time,

i.e. $\text{Ido}[i,\tau]$ has the possible *smallest* value.

All the three variants are clearly greedy ones, but in the great level of lack of information, we have no other simple idea for solving PMS.

## 3 Running experiments

So far **we have tested the above algorithms on** large size **input datasets, in which each block contained** *only one* **task**, i.e. when the scheduling has *no* possibility for solving tasks parallel. Further research for the solution of the general problem is in progress, will be summarized in [5] .

Let us start with a short illustrative example:

**Input** (`Ls4.dat`):

```
 8  = t = number of Task-types
 3  = H = number of Machines
 2  = K = number of tasks per Jobs
10  = L = number of Jobs
------------------------------------+
 2  5  1  9  5  6  7  2  / Machine 1.|  = I[h,τ]
 3  4  8  7  4  1  3  1  / Machine 2.|   table of times
 4  6  9  8  3  4  6  3  / Machine 3.|   neccessary for
------------------------------------+   M_h for T_τ
 1  2   / Job  1.
 3  4   / Job  2.
 1  3   / Job  3.
 2  3   / Job  4.
 3  3   / Job  5.
 2  6   / Job  6.
 5  2   / Job  7.
 3  5   / Job  8.
 1  6   / Job  9.
 4  8   / Job 10.
 0      / END
```

The solution of variant )3( can be seen on Figure 1.

Before any running experiments or theoretical investigations one might have the following *prejudices*  (only few of them were justified during the runs):

" Variant  )3(  may give the *best* scheduling solution but it runs slowest (since it is the most precise?) " ,

" Variant  )4(  may be the *quickest* but it provides bad solution  (since it deals with the starting time only?) ",

" Variant  )5(   may be *quickest* but it provides not so good + not so bad solution  (since it may give many tasks to some $M_{ho}$ ) ".

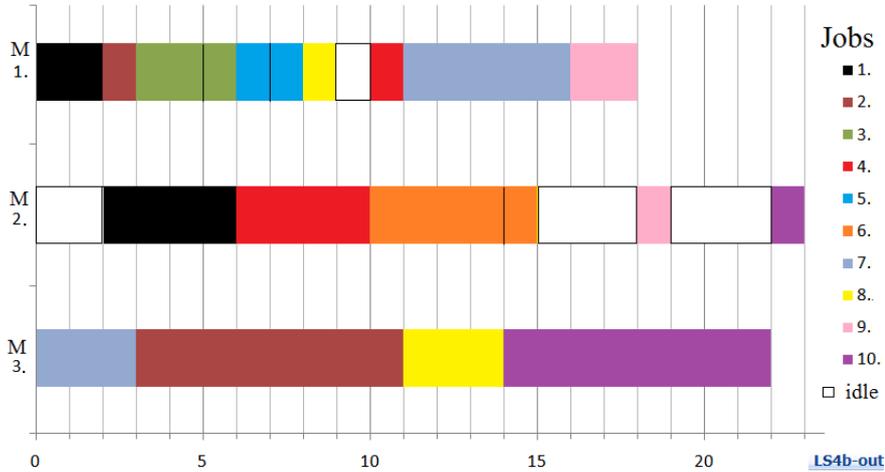

*Figure 1:* A solution of the example `Ls4.dat`

We tested the variants (3) - (5) with *many* (more than 600) *large*, randomly (uniformly) generated datasets, each of them containing 500-1000 jobs. We used a personal computer with Intel® Core™ **Quad** CPU Q6600 @ 2.40GHz, 4GB RAM, Windows 7 and Delphi 7 language.

The summary of one of the runs can be seen in [4], which includes and comparises schedulig- and running times, too.

Let us first serve the details of running of *medium size* datasets. N stands for the number of the datasets. Determining

$$1 \leq t \leq 100,\ 1 \leq m \leq 100,\ 1 \leq \mathbf{I}(h,\tau) \leq 100,\ 1 \leq L \leq 200,\ 1 \leq K \leq 20,\ N=500, \quad (6)$$

the average *running time* (for each dataset) were some minues, the *sizes* of the ouput files were 10kb - 1Mb separately, 100Mb total. These data are valid for all the three variants.

There were some differences in *running times*:
variant (5) is highly faster than (3), in detail :
  in **14%** cases  $0.5 * (3) < (5) \leq 0.9 * (3)$
  in **75**% cases         $(5) < 0.5 * (3)$

However, the resulted *scheduling times* (solutions) were totally different (see [4] for details):

variant (4) has extremaly bad solutions in almost all cases,
variant (3) has **the best** solution with few exceptions:
  in **6%** cases  (5) is better, than (3),

   in **17%** cases  )5(  is as good as  )3( ,  
variant  )5(  has **not so bad** solutions:  
   in **29%** cases  **1.0 \*** )3(  <  )5(  <  **1.2 \*** )3(  
   in **42%** cases  **1.2 \*** )3(  ≤  )5(  <  **2.0 \*** )3(  

As we observed, the above data *slightly depend* upon the size of the problem.

Now, let us turn to the experiment of *large* datasets. Relaying on the bad experiences in the previous subsection, we *excluded* variant  )4(  from our further experiments. Our settings were:

$$1 \leq t, m, \mathbf{I}(h,\tau), L, K \leq 1000, \quad N=650. \qquad (7)$$

The average *running times*  (for all the N=650 datasets, total) :  
   variant  )3(  runned for  81 hours (!),  
   variant  )5(  runned for  2 hours,  
i.e. the difference is large: variant  )5(  is tremendously faster than )3( :  
   in **23%** cases  **0.5 \*** )3(  <  )5(  ≤  **0.50 \*** )3( ,  
   in **75%** cases  )5(  <  **0.05 \*** )3(  !  

The *sizes* of ouput files (separately) were between 1Mb and 50Mb, 4Gb (!) total for both these variants.

The rate of resulted *scheduling times* showed oposite direction:

variant )3( has **always the best** solution, )5( is **fairly bad**  
   in **11%** cases  **1.0 \*** )3(  <  )5(  <  **1.2 \*** )3( ,  
   in **29%** cases  **1.2 \*** )3(  ≤  )5(  <  **2.0 \*** )3( ,  
   in **33%** cases  **2.0 \*** )3(  ≤  )5(  <  **3.0 \*** )3( ,  
   in **25%** cases  **3.0 \*** )3(  ≤  )5( .  

## Conclusions

In general, but especially in large size datasets, version )5( was exponentially faster than version )3( , but considering the resulted schedulings, version )5( gave worse results than version )3( , in *moderate manner*. This shows again the old dilemma: "shorter running time" *versus* "better results " !

We have *no* comparison with the *absolute* (offline) optimum, but )3( **might be optimal** (in the offline sense) **in many cases**, since there are *very few idle time* (pause) of the machines, and they finish almost at the same "moment".

**Acknowledgements**

We say grateful thanks for the following sponsors.

Research was supported by the Fund TÁMOP and National Research Center for Development and Market, Introduction of Avanced Information and Communication Technologies, TÁMOP 4.2.2.C-11/1/KONV-2012-0004, II./4.

Publishing of this Conference Proceeding is supported by the Sapientia University.